\documentclass[sigconf]{acmart}   
\usepackage{subcaption}
\usepackage{bm}
\usepackage[ruled,vlined]{algorithm2e}
\usepackage{algorithmic}
\usepackage[inline]{enumitem}
\usepackage{mathtools}
\usepackage{amsmath}
\usepackage{footnote}
\usepackage{tablefootnote}
\usepackage{longtable}
\usepackage{multirow}
\usepackage{lipsum}
\usepackage{calc}
\usepackage{colortbl}
\usepackage{float}
\usepackage[flushleft]{threeparttable} 
\usepackage{booktabs,caption}
\definecolor{lightgray}{gray}{0.9}

\acmConference[]{}{}{}

\renewcommand\footnotetextcopyrightpermission[1]{} 
\begin{document}

\title{Probabilistic Model of Narratives Over Topical Trends in \\ Social Media: A Discrete Time Model}

\author{Toktam A. Oghaz$^{1}$, Ece C. Mutlu$^{2, *}$, Jasser Jasser$^{1, *}$, Niloofar Yousefi$^{2}$, 
Ivan Garibay$^{1,2}$}
\email{{toktam@cs, ece.mutlu@, jasser.jasser@, niloofar.yousefi@, igaribay@}.ucf.edu}
\thanks{*Equal contribution.}
\affiliation{%
  \institution{$^1$Department of Computer Science,\\
  $^2$Department of Industrial Engineering,\\
  University of Central Florida}
}





\renewcommand{\shortauthors}{ }

\begin{abstract}
Online social media platforms are turning into the prime source of news and narratives about worldwide events. However,a systematic summarization-based narrative extraction that can facilitate communicating the main underlying events is lacking. To address this issue, we propose a novel event-based narrative summary extraction framework. Our proposed framework is designed as a probabilistic topic model, with categorical time distribution, followed by extractive text summarization. Our topic model identifies topics' recurrence over time with a varying time resolution. This framework not only captures the topic distributions from the data, but also approximates the user activity fluctuations over time. Furthermore, we define significance-dispersity trade-off (SDT) as a comparison measure to identify the topic with the highest lifetime attractiveness in a timestamped corpus. We evaluate our model on a large corpus of Twitter data, including more than one million tweets in the domain of the disinformation campaigns conducted against the White Helmets of Syria. Our results indicate that the proposed framework is effective in identifying topical trends, as well as extracting narrative summaries from text corpus with timestamped data.

\end{abstract}




\keywords{Topic Modeling; Graphical Models; Narrative\footnote{In this study, we have used the terms narrative and story interchangeably.} Extraction; Topic Detection and Tracking; Extractive Text Summarization }

\maketitle

\section{Introduction}\label{introduction}
Social media and microblogging platforms, such as Twitter and Facebook, are becoming the primary sources of real-time content  regarding ongoing socio-political events, such as United States Presidential Election in 2016 \cite{enli2017twitter}, and natural and man-made emergencies, such as COVID-19 pandemic  in 2020 \cite{cinelli2020covid}. However, without the appropriate tools, the massive textual data from these platforms makes it extremely challenging to obtain relevant information on significant events, distinguish between high-quality and unreliable content \cite{hassan2017claimbuster}, or identify the opinions within a polarized domain \cite{garibay2019polarization}. 

The challenges mentioned above have been studied from different aspects related to topic detection and tracking within the field of Natural Language Processing (NLP). Researchers have developed automatic document summarization tools and techniques, which intend to provide concise and fluent summaries over a large corpus of textual data \cite{maryam2018twitter}. 
Preserving the key information in the summary and producing summaries that are comparable to human-created narratives are the primary goals of the extractive and abstractive approaches for automatic text summarization \cite{allahyari2017text}. News websites are a prime example of such techniques, where automatic text summarization algorithms are applied to generate news headlines and titles from the news content \cite{tan2017neural}. 
The shortage of labeled data for text analysis has encouraged researchers to develop novel unsupervised algorithms that consider co-occurrence of words in documents as well as emerging new techniques such as exploiting an additional source of information similar to Wikipedia knowledge-based topic models \cite{xu2017incorporating, yao2016concept}. Additionally, unsupervised learning enables training general-purpose systems that can be used for a variety of tasks and applications as strong classifiers \cite{burkhardt2019survey}. In this regard, statistical models of co-occurrence such as Latent Dirichlet Allocation (LDA) \cite{blei2003latent}, discover the relevant structure and co-occurrence dependencies of words within a collection of documents to capture the distribution of topic latent variable from the data. Although an abundant timestamped textual data, particularly from social media platforms and news reports are available for analysis, the changes in the distribution of data over time have been neglected in most of the topic mining algorithms proposed in the literature \cite{wang2006topics}. For instance, time-series analysis on datasets over the events relative to 2012 US presidential election suggests that modeling topics and extracting summaries without considering the text-time relationship lead to missing the rise and fall of topics over time, the changes in terms of correlations, and the emergence of new topics \cite{gad2015themedelta}. 

Although continuous-time topic models such as \cite{wang2006topics} have been proposed in the literature, topical models with continuous-time distribution cannot model many modes in time, which leads to deficiency in modeling the fluctuations. Additionally, continuous-time models suffer from instability problems in the case of analyzing a multimodal dataset that is sparse in time. 

In this paper, we propose a probabilistic model of topics over time with categorical time distribution to detect topical recurrence, designed as an LDA-based generative model. 
To achieve probabilistic modeling of narratives over topical trends, we incorporate the components of narratives including named-entities and temporal-causal coherence between events into our topical model.  
We believe that what differentiates a narrative model\footnote{In this paper, we refer to event-based topic modeling over topical trends as narrative modeling.} from topic analysis and summarization approaches is the ability to extract relevant sequences of text relative to the corresponding series of events associated with the same topic over time. 
Accordingly, our proposed narrative framework integrates unsupervised topic mining with extractive text summarization for narrative identification and summary extraction. 
We compare the identified narratives by our model with the topics identified by Latent Dirichlet Allocation (LDA) \cite{blei2003latent} and Topics over Time (TOT) \cite{wang2006topics}. This comparison includes presenting numerical results and analysis for a large corpus of more than one million tweets in the domain of disinformation campaigns conducted against the White Helmets of Syria. The collected dataset contains tweets spanning 13 months within the years 2018 and 2019. Our results provide evidence that our proposed method is effective in identifying topical trends within a timestamped data. Furthermore, we define a novel metric called significance-dispersity trade-off (SDT) in order to compare and identify topics with higher lifetime attractiveness in timestamped data. Finally, we demonstrate that our proposed model discovers time localized topics over events that approximates the distribution of user activities on social media platforms.

The remaining of this paper is organized as follows: First, an overview of the related works is provided in Section \ref{sec:related_work}. In Section \ref{sec:proposed}, we provide a detailed explanation of our proposed method followed by the experimental setup and results.
Finally, in Section \ref{sec:conclusion} we conclude the paper and discuss future directions.

\section{Background and Related Work}\label{sec:related_work}
In this section, we first provide a background on narrative analysis and how literature has investigated stories in social media. Then, we present an overview of topic modeling and text summarization. 

\subsection{Narrative analysis}
Narratives can be found in all day-to-day activities. The fields of research on narrative analysis include narrative representation, coherence and structure of narratives, and the strategies, aim, and functionality of storytelling \cite{mishler1995models}. From a computational perspective, narratives may relate to topic mining, text summarization,  machine translation \cite{vargas2017narrative}, and graph visualization. The later can be achieved via using directed acyclic graphs (DAGs) to demonstrate relationships over the network of entities \cite{glavavs2014hieve}. Narrative summaries can be constructed from an ordered chain of individual events with causality relationships amongst events, appeared within a specific topic \cite{jans2012skip}. The narrative sequence may report fluctuations over time relative to the underlying events. Additionally, the story-like interpretation of the text is a must to imply a narrative \cite{page2013seriality}. 


Since social media have been admitted as a component of today's society, many studies have investigated narratives in social media content \cite{georgakopoulou201717, page2013seriality, veel2018make}. These Narratives contain small auto-biographies that have been developed in personal profiles and cover trivial everyday life events. Other types of narratives appearing in social media platforms consist of breaking news and 
long stories of past events \cite{page2013seriality}.
Some types of narratives, such as breaking news, result in the emergence of other narratives related to the predictions or projections of events in near future \cite{georgakopoulou201717}. 
These literature view social media conversation cascades as 
stories that are co-constructed by the tellers and their audience, and are circulating amongst the public within and across social media platforms. 
Moreover, the events have been considered as the causes of online user activity that can be identified via activity fluctuations over time \cite{ansah2020leveraging, page2013seriality}. 
Developing appropriate tools for social media narrative analysis
can facilitate communicating the main ideas regarding the events in large data. 

\subsection{Topic Mining and Text Summarization}
As social media activities generate abundant timestamped multimodal data, many studies such as \cite{chua2013automatic} have presented algorithms to discover the topics and develop descriptive summaries over social media events. 
probabilistic models to discover word patterns that reflect the underlying topics in a set of document collections \cite{alghamdi2015survey}. The most commonly used approach to topic modeling is Latent Dirichlet Allocation (LDA) \cite{jelodar2019latent}. LDA is a generative probabilistic model with a hierarchical Bayesian network structure that can be used for a variety of applications with discrete data, including text corpora. Using LDA for topic mining, a document is a bag-of-words that has a mixture of latent topics \cite{blei2003latent}.
Many advanced topic modeling approaches have been derived from LDA, including Hierarchical Topic Models \cite{griffiths2004hierarchical, glavavs2014hieve} that learn and organize the topics into a hierarchy to address a super-sub topic relationship. This approach is well-suited for analyzing social media and news stories that contain rich data over a series of real-world events \cite{srijith2017sub}. Topic models over time with continuous-time distribution \cite{blei2006dynamic} and dynamic topic models \cite{wang2006topics} intend to capture the rise and falls of topics within a time range. However, continuous-time topic models, such as beta or normal time distribution, cannot model many modes in time. Furthermore, the smooth time distribution over topics does not allow recognizing distinct topical events in the timestamped dataset, where topical events reflect the event-based topic activity fluctuations over time. 

Topic modeling and summarization of social media data is challenging as a result of certain restrictions, such as the maximum number of characters allowed on the Twitter platform. As short-text or microblogs have low word co-occurrence and contextual information, models designed for short-text topic analysis and summarization may obtain context information with short-text aggregation to enrich the relevant context before further analysis \cite{quan2015short}. 

Document summarization techniques are generally categorized into abstractive and generative text summarization models. Herein, we consider extractive text summarization methods. Several algorithms for extractive text summarization have been proposed in the literature that assign a salient score to sentences \cite{dutta2019summarizing}. To summarize a text corpus with short text, \cite{shou2013sumblr} presents an automatic summarization algorithm with topic clustering, cluster ranking and assigning scores to the intermediate features, and sentence extraction. 
Some other approaches, particularly for the Twitter data include aggregating tweets by hashtags or conversation cascades \cite{quan2015short, torres2020seq2seq}, and obtaining summaries for a targeted event of interest as one or a set of tweets that are representative of the topics \cite{chua2013automatic}. 

Additionally, neural network-based summarization models \cite{ren2017leveraging, narayan2018ranking}, commonly with an encoder-decoder architecture, leverage attention mechanism for contextual information among sentences or ROUGE evaluation metric to identify discriminative features for sentence ranking and summarization. However, these architectures require labeled datasets and might not apply to short-text. Text summarization with compression using neural networks is proposed by \cite{xu2019neural} which applies joint extraction and syntactic compression to rank compressed summaries with a neural network. 
Our focus in the present work is on probabilistic topic modeling and extractive text summarization to provide descriptive narratives for the underlying events that occurred over a period of time.







\section{Methodology}\label{sec:proposed}
In this section we explain our narrative framework. The framework comprises of 2 steps: 
\begin{enumerate*}[label=\Roman*.]
    \item Narrative modeling based on topic identification over time and
    \item extractive summarization from the identified narratives. 
\end{enumerate*}
To discover the narratives over topical events, first, we use our discrete-time generative narrative model as an unsupervised learning algorithm to learn distribution of textual contents from daily conversation cascades. Then, we extract narrative summaries over topical events from sentences in the time categories. This is achieved by sampling from the identified distribution of narratives and perform sentence ranking. 
Narrative modeling and summarization steps are explained below in separate subsections.

\subsection{Narrative Modeling}
To model narratives, we design our topic model such that the discovered topics present a series of timely ordered topical events. Accordingly, the topical events deliver a narrative covering distinct events over the same topic. In this regard, we present Narratives Over Categorical time (NOC), a novel probabilistic topic model that discovers topics based on both word co-occurrence and temporal information to present a narrative of events. According to the topic-time relationship explained above, we refer to the topics or narratives, topical events as events, and the extracted timely ordered sentences of documents with high probability of belonging to each event as the extracted narrative summary. To fully comply with the definition of narrative, we assume a causality relation between the conversation cascades in social media. However, we do not investigate the causality relation across the conversation cascades or named-entities. 

The differences between our Narrative model with dynamic topic models \cite{blei2006dynamic}, topic models with continuous time distribution \cite{wang2006topics}, and hierarchical topic models \cite{griffiths2004hierarchical, pujara2012large} include: not filtering the data for an specific event, imposing sharp transition for topic-time changes with time slicing, discovering topical events without scalability and sparsity issues, allowing multimodal topic distribution in time as a result of categorical time distribution, and selecting an appropriate slicing size such that distinct topical events be recognizable. 
Additionally, categorical time distribution enables discovering topical events with varying time resolution, for instance, weekly, biweekly, and monthly. 

Time discretization brings the question of selecting the appropriate slicing size or the number of categories that depends on the characteristics of the dataset under study. On the contrary, topical models with continuous time distribution cannot model many modes in time. Additionally, continuous time models such as \cite{wang2006topics} suffer from instability problem if the dataset is multimodal and sparse in time. Furthermore, categorical time enables discovering topic recurrence  which results in identifying topical events related to distinct narrative activities, which is of our interest in this paper. Narrative activities in social media refer to the amount of textual content that is circulating in online platforms over time, corresponding to a specific topic. 



\begin{figure}[t]
    \centering 
    \includegraphics[width=.4\textwidth]{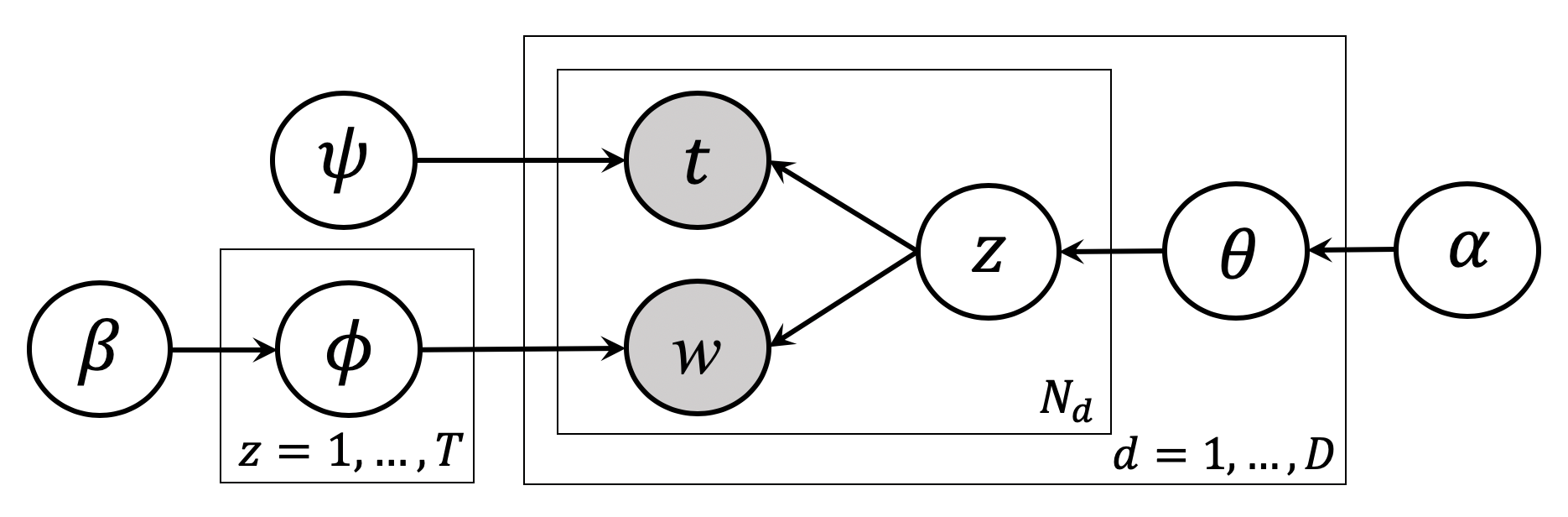}
    \caption[  ]
    {\small The graphical model for NOC with Gibbs sampling.
    } 
    \label{fig:graphm}
\end{figure}

The generative process in NOC, models timestamps and words per documents using Gibbs sampling which is a Markov Chain Monte Carlo (MCMC) algorithm. The graphical model of NOC is illustrated in Figure \ref{fig:graphm}. As can be seen from the graphical model, the posterior distribution of topics is dependent on both text and time modalities. This generative procedure can be described as follows:

\begin{enumerate}[label=\Roman*.,leftmargin=2\parindent]
   \setlength\parindent{24pt} \item For each topic $z$, draw $T$ multinomials $\phi_z$ from a Dirichlet prior $\beta$;

    \item For each document $d$, draw a multinomial $\theta_d$ from a Dirichlet prior $\alpha$;

    \item For each word $w_{di}$ in $d$: 
      \begin{enumerate}[leftmargin=2\parindent]
        \item draw a topic $z_{di}$ from multinomial $\theta_d$;
        \item draw a word $w_{di}$ from multinomial $\phi_{z_{di}}$; 
        \item draw a timestamp $t_{di}$ from categorical $\psi_{z_{di}}$;
      \end{enumerate}

\end{enumerate}
\noindent
where the timestamps $t_{di}$ for words $w_{di}$ in each document $d$ are identical. The list of symbols and their descriptions can be found in table \ref{tab:symbols}. 
The model parameterization is as below:

\begin{equation}
\small
\begin{aligned}
 \theta_{d}|\alpha &\sim ~\text{Discrete}(\alpha) \\
 \phi_z | \beta &\sim ~\text{Discrete}(\beta) \\
 z_{di}|\theta_d &\sim ~\operatorname{Multinomial}(\theta_d) \\
 w_{di}|\phi_{z_{di}} &\sim ~\operatorname{Multinomial}(\phi_{z_{di}}) \\
 t_{di}|\psi_{z_{di}} &\sim ~\operatorname{Categorical}(\psi_{z_{di}}) \\
\label{eq:gen_model}
\end{aligned}
\end{equation}

\begin{table}[t]
  \caption{Symbols and definitions used in this paper}
  \centering
  \begin{tabular}{ p{6.8cm} c}
    \toprule
    Variable Descriptions   & Symbol \\
    \midrule
    Number of topics                                        &    $T$ \\
    Number of documents                                     &    $D$ \\
    Number of unique words                                  &    $V$ \\
    Number of word tokens in document d                     &    $N_d$ \\
    Multinomial distribution of topics for document d   &    $\theta_d$ \\
    Multinomial distribution of words for topic z       &    $\phi_z$ \\
    Categorical distribution of time for topic z        &    $\psi_z$ \\
    Topic of the $i$th token in document d              &    $z_{di}$ \\
    $i$th word token in document d                      &    $w_{di}$ \\
    Timestamp for $i$th word token in document d        &    $t_{di}$ \\
    Time category for timestamp associated with a token  &    $b_k$ \\
    Entropy of topic z                                    & $H_z$ \\
    $j$th sentence of document d                        & $s_{dj}$  \\
    
    \bottomrule
  \end{tabular}
  \label{tab:symbols}
\end{table}

\begin{figure*}
    \centering 
    \includegraphics[width=\textwidth]{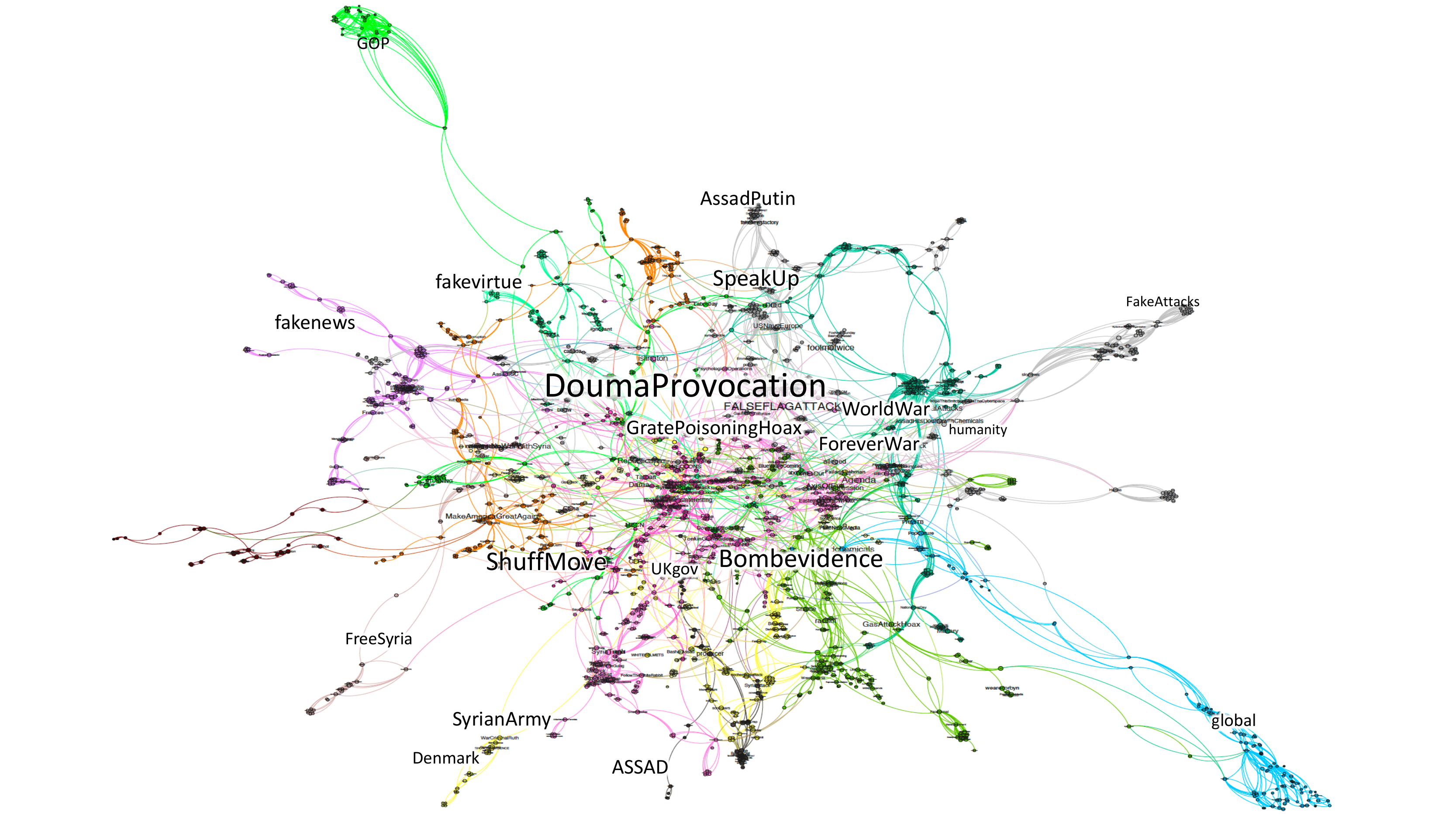}
    \caption[  ]
    {\small The hashtag co-occurrence graph for twitter dataset on the domain of White Helmets of Syria for a period of 13 month, from 2018 to 2019. This graph represents a down-sampled version of the hashtag co-occurrence in this data for the sake of visualization.
    } 
    \label{fig:hashtagsco}
\end{figure*}

In this model, Gibbs sampling provides an approximate inference instead if exact inference. To calculate the probability of topic assignment to word $w_{di}$, we first need to calculate the joint probability of the dataset as $\mathbb{P} (\boldsymbol{z_{di}, w_{di}, t_{di}} | w_{-di}, t_{-di}, z_{-di}, \alpha, \beta, \psi)$ and use chain rule to derive the probability of $\mathbb{P} (\boldsymbol{z_{di}} | \boldsymbol{w},\boldsymbol{t}, \boldsymbol{z_{-di}}, \alpha, \beta, \psi )$ as below, where $-di$ subscripts refers to all tokens except $w_{di}$:

\begin{equation}
\begin{aligned}
\mathbb{P} (z_{di} | \boldsymbol{w},\boldsymbol{t}, \boldsymbol{z_{-di}}, \alpha, \beta, \psi ) 
& \propto 
(m_{dz_{di}} + \alpha_{z_{di}} - 1) \\
\times 
\frac{n_{z_{di}w_{di}} + \beta_{w_{di}} - 1}{\sum_{v = 1}^{V} (n_{z_{di}v} + \beta_v) - 1} & \;
p(t_{z_{di}} \in b_k)
\label{eq:gibbs}
\end{aligned}
\end{equation}

\noindent
where $n_{zv}$ refers to the number of words $v$ assigned to topic $z$, $m_{dz}$ refers to the number of word tokens in document $d$ that are assigned to topic $z$, and $b_k$ represents the $k$th time slice. The details on the Gibbs sampling derivation can be found in the Appendix section. 
After each iteration of Gibbs sampling, we update the probability of $p(t_{z_{di}} \in b_k)$ as follows:

\begin{equation}
\begin{aligned}
p(t_{z_{di}} \in b_k) = \frac{1}{K}\sum_{k = 1}^{K} \mathbb{I}(t_{z_{di}} \in b_k) 
\end{aligned}
\end{equation}
where $\mathbb{I}(.)$ is equal to 1 when $t_{z_{di}} \in b_k$, and 0 otherwise. 

In this paper, we report results with bi-weekly categorical time resolution. To determine the values for hyper-parameters $\alpha$ and $\beta$ and to investigate the sensitivity of the model to these values, we repeated our experiment with symmetric Dirichlet distributions using values $\alpha \in [0.1, 0.5, 1]$, $\beta \in [0.01, 0.1, 0.5, 0.8, 1]$. We observed that the model did not show significant sensitivity to the values of these hyper-parameters. Thus, we fix $\alpha = 1$ and $\beta = 0.5$, both as symmetric Dirichlet distributions. We initialize the hyperparameter $\psi$ in 2 ways for comparison: 
\begin{enumerate*}[label={\Roman*.}]
\item random initialization (model referred as $NOC_R$); and 
\item based on the probability of user activity per time category, illustrated in Figure \ref{fig:activity}, (model referred as $NOC_A$).
\end{enumerate*}

To estimate the number of topics for our experiments, we first visualize the tweets' hashtag co-occurrence graph. We measure the graph modularity to examine the structure of the communities in this graph. We observe the highest modularity score of 0.41 using modularity resolution equal to 0.85. Figure \ref{fig:hashtagsco} illustrates a downsample version of this graph, where each color represents a modularity class. The edges of the graph are weighted according to the number of hashtags' co-occurrence in the document collection. Our modularity analysis suggests that few distinct hashtag communities exist. Additionally, the dataset under study contains tweets associated with a single domain. As a result, we assume the number of topics to be relatively low. To choose an appropriate number of topics, we repeated LDA with the number of topics as $T \in [4, \dots, 20]$ with increments of size 1. We evaluated the $c_v$ coherence of topics identified by LDA and observed the highest coherence score for $T = 5$ and $T = 5$, respectively. Thus, we report our experimental results using these values.

\subsection{Narrative Summary Extraction}
We employ the discovered probabilities of topics over documents, $\theta$, probabilities of words per topic, $\phi$, and probabilities of topics per time category, $\psi$ to perform sentence ranking. This ranking allows extracting the sentences with the higher scores of belonging to each topic. This is achieved via performing weighted sampling on the collection of documents based on the probabilities of topics per time category $\psi$ and draw $D$ documents from $\theta$. The weighted sampling leads to drawing more documents from the time categories $b_k$ with a higher $\psi$ as this time slices contain more documents related to the topic $z$. 
Each document contains a sequence of sentences $(s_1, s_2, \dots, s_J) \in d$ from the aggregated conversation cascades per day. Information on the aggregation of conversation cascades and document preparation can be found in section \ref{sec:data}. 

Since the social media narrative activity over a topic evolves from the circulation of identical or similar textual content in the platform, the content involves significant similarity. For instance, the Twitter conversation cascades include replies, quotes, and comments, where replies and quotes duplicate the textual content. Therefore, we applied Jaro-Winkler distance over the timely ordered sentences and dismissed the sentences with similarity above 70\%, while keeping the longest sentence. After removing redundant text as described earlier, we calculate the probability of each sentence $s_j$ by measuring the sum of the probabilities of topics for words $w_{di} \in s_j$. Then, we select the sentences
with the highest accumulative probability of words $w$ per topic $z$. 
Summary coherence was induced as suggested in \cite{barzilay2001sentence} by ordering the extracted sentences according to their timestamps such that the oldest sentences appear first. Table \ref{tab:summary} in the Appendix section contains the extracted narrative summaries for 5 topics for a sample run. 




\section{Experiments and Results}
\begin{figure}[htp]
\begin{subfigure}{\columnwidth}
  \centering
  \includegraphics[width=\linewidth]{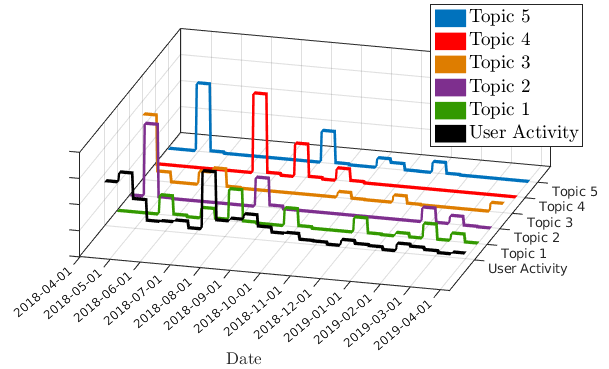}  
  \caption{}
  \label{fig:alltopics}
\end{subfigure}
\begin{subfigure}{\columnwidth}
  \centering
  \includegraphics[width=\linewidth]{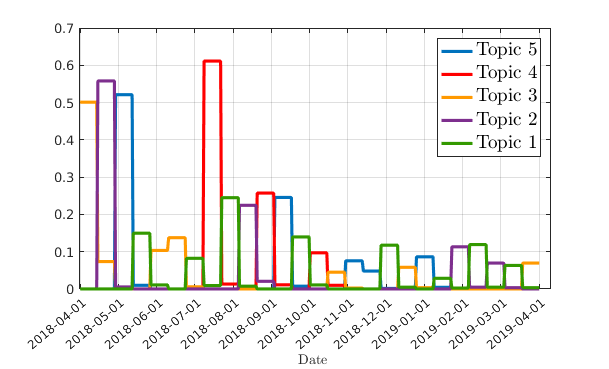}  
  \caption{}
  \label{fig:collapsed}
\end{subfigure}
\begin{subfigure}{\columnwidth}
  \centering
  \includegraphics[width=\linewidth]{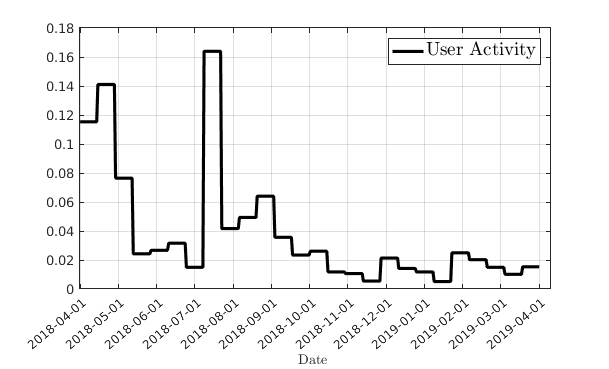}  
  \caption{}
  \label{fig:activity}
\end{subfigure}

\caption{
The distribution of extracted topics and user activity over time: a) The distribution of user activity over time is depicted by black, followed by the 5 distribution of the extracted topics, using $NOC_R$; b) The collapsed distribution of the 5 extracted topics; c) The distribution of user activity in time. The results suggest that the distribution of extracted topics approximate the distribution of user activity over the same time period.}
\label{fig:topicsdis}
\end{figure}

As mentioned earlier, the discovered topics by NOC present a series of timely ordered topical events. Thus, the topical events deliver a narrative covering distinct social media events over the same topic. 
Figure \ref{fig:topicsdis} demonstrates the generated narrative distributions with NOC, where the hyperparameter $\psi$ was randomly initialized (referred to as NOC\textsubscript{R}). This figure represents that the identified narratives by our model are distinct from each other and the collapsed distribution of all narratives approximates the distribution of social media user activity over time. 

The identified narratives can be evaluated using effective evaluation metrics for topic models. Accordingly, we calculate pointwise mutual information \cite{newman2010automatic} to measure the coherence of a topic $z$ as follows: 

\begin{equation}
\begin{aligned}
Coh_{z} = \frac{2}{K (K - 1)} \sum_{j < k \leqslant K} log \frac{p(w_j, w_k)}{p(w_j)p(w_k)}
\label{eq:coherence}
\end{aligned}
\end{equation}
where $K$ is the number of most probable words for each narrative, $p(w_j)$ and $p(w_k)$ refer to the probabilities of occurrence for words $w_j$ and $w_k$, and $p(w_j, w_k)$ represents the probability of co-occurrence for the two words in the collection of documents. 

We compare our model with LDA and TOT\cite{wang2006topics}, where TOT is a probabilistic topic model over time with Beta distribution for time. Table \ref{tab:coherence} displays the average coherence score measured across the discovered topics by LDA, TOT, and NOC. For NOC, we investigate initializing the parameter $\psi$ with random and user activity-based initialization, referred as NOC\textsubscript{R} and NOC\textsubscript{A}, respectively. We considere $K = 500$ most probable words from each topic. This comparison suggests that the narratives identified by NOC are more coherent than the identified topics by LDA, with an improvement in coherence of about 35\%. The observed improvement comparing with TOT was about 27\%. Additionally, initializing the hyperparameter $\psi$ in NOC using the distribution of user activity improves the narrative coherence by about 3\%. 

\begin{table}[t]
  \caption{The comparison of coherence scores for 4 models:}
  \centering
  \begin{tabular}{c| c c c c}
    \toprule
    
    \textbf{\emph{Model}}  &    \textbf{\emph{LDA}} &  \textbf{\emph{TOT
    \tablefootnote{We used the implementation available on \url{https://github.com/ahmaurya/topics_over_time.}}
    }}  &
    \textbf{\emph{NOC\textsubscript{R}}}   &   \textbf{\emph{NOC\textsubscript{A}}} \\
    \midrule
    \textbf{\emph{T = 5}}  &    5.980  &    6.36  & 7.95       &  \textbf{8.23} \\
     \textbf{\emph{T = 6}}  &    5.546  &   5.99   & 7.75       &  \textbf{7.98} \\
    
    \bottomrule
  \end{tabular}
  \label{tab:coherence}
\end{table}

\subsection{Proposed Evaluation Metric}
The topic attractiveness to social media users can be  investigated as a measure of the length of conversation cascades, the number of initiated textual content, and the number of unique users performing an activity relative to the underlying topic.
The user activity fluctuations for timestamped data may contain activity bursts that are illustrative of significant events.  
Similarly, the generation and propagation of textual content within an online platform can illustrate the narrative activity relative to the events over time, where a burst represents a significant narrative activity.  
Additionally, the recurrence of a topic can be considered as an attractiveness measure for the associated topic. 

In this regard, we propose the significance-dispersity trade-off (SDT) metric to compare the identified narratives against each-other. SDT measures the lifetime attractiveness of the identified narratives based on the distribution of narratives over topical events. The proposed metric quantifies the significance of the narrative activities and recurrence of a topic via employing the Shannon entropy for the discovered narrative distributions. The intuition behind the SDT score is that the value of the entropy is maximum when the probability distribution is uniform. On the contrary, this value is minimum if the distribution is delta function. This is visualized in Figure \ref{fig:sdt} in the Appendix section. 

We define dispersity of a categorical time topic distribution as a measure of the dispersion of the time categories. Based on this definition, SDT score of topic $z$ can be obtained as:
\begin{equation}
    SDT_z = H^{\gamma} ( H_{max} - H)^{1 - \gamma},
\end{equation}
where H is the Shannon entropy for the categorical distribution of time for topic $z$:

\begin{table}[t]
  \caption{The comparison of SDT scores for 5 narratives:}
  \centering
  \begin{tabular}{c| c c c c c}
    \toprule
    
    \textbf{\emph{Narrative}}  &    \textbf{\emph{T\textsubscript{1}}} & 
    \textbf{\emph{T\textsubscript{2}}} &
    \textbf{\emph{T\textsubscript{3}}} &
    \textbf{\emph{T\textsubscript{4}}} &
    \textbf{\emph{T\textsubscript{5}}} \\
    
    \midrule
    $\gamma = 0$  & 1.59  & 2.90  &  2.39 & \textbf{3.21} & 2.75 \\
    $\gamma = 0.4$  & 2.08  & \textbf{2.40}  &  2.36 & 2.36 & \textbf{2.40} \\
    $\gamma = 0.7 $  & \textbf{2.54} & 2.08  &  2.33 & 1.87 & 2.16 \\
    $\gamma = 1 $  & \textbf{3.11} & 1.80  &  2.31 & 1.49 & 1.95 \\
    
    \bottomrule
    User Activity  & \textbf{353,280} & 317,686  &  244,674 & 247,895 & 175,343 \\
    \bottomrule
  \end{tabular}
  \label{tab:sdt}
\end{table}

\begin{equation}
   \hfill H_z = - \sum_{k}^K p_z log_2 p_z, 
\end{equation}
\begin{equation*}
H_{max} = log_2 (K),
\end{equation*}
and $K$ refers to the number of time slices in the distribution. 
We assume that social media topics with high lifetime attractiveness are significant and recurrent. However the probability distribution imposes a trade-off on the two. 
The parameter $\alpha$ provides a weighted geometric mean of $H$ and $H_{max} - H$ that enables promoting either significance or recurrence, dependent on the application under study. A larger value of parameter $\alpha$ promotes dispersity for SDT score, and a smaller amount of this parameter promotes mode significant. The bounds for the SDT score are:
\begin{equation}
  SDT_i =
    \begin{cases}
      0 &               \text{if $H = 0$ \& $\gamma != 0$}\\
      0 &         \text{if $H = H_{max}$ \& $\gamma != 1$} \\
      \gamma^{\gamma} (1 - \gamma)^{1 - \gamma} H_{max}   & \text{if $H = \gamma H_{max}$ \& $0 < \gamma$ < 1}
    \end{cases} 
\end{equation}
where $H = 0$ occurs when the distribution under study is uniform, and $H = H_{max}$ relates to delta distribution. 
Since the time categorical distribution of our narrative model allows many modes in time, recurrent narratives can be identified. Additionally, the narrative activity fluctuations can be modeled using categorical time distribution in topic analysis.
Table \ref{tab:sdt} provides a comparison for the SDT scores measured for the 5 identified narratives, using varying values of $\alpha$. The illustration of the distribution of the extracted narratives can be seen in Figure \ref{fig:alltopics}. We can clearly see in this figure that narratives 1 and 3 have the highest dispersity. On the contrary, narratives 4 and 2 have the highest significance. We compare $SDT_i$ for narrative $i$ with the number of user activity associated with narrative $z$. The results suggest that SDT score can be used to identify the narrative with higher lifetime attractiveness in a timestamped dataset. In our experiments, this is achieved for topic 1 when the value of $\gamma$ is greater than or equal to 0.7. As it can be seen, this topic is associated with the highest user activity count, reported in the same table. 







\subsection{Dataset Description and Pre-processing}
\label{sec:data}
To analyze topical events and provide narratives, we investigate the Twitter dataset on the domain of White Helmets of Syria over a period of 13 month from April 2018 to April 2019. This dataset was provided to us by Leidos Inc\footnote[1]{\url{https://www.leidos.com/}} as part of the Computational Simulation of Online Social Behavior (SocialSim)\footnote[2]{\url{https://www.darpa.mil/program/computational-simulation-of-online-social-behavior}} program initiated by the Defense Advanced Research Projects Agency (DARPA). We analyze more than 1,052,000 tweets from April 2018 to April 2019.   

To prepare the model inputs, we filter the tweets from non-English text. Then, we clean up the data by removing usernames, short URLs, as well as emoticons. Additionally, we remove the stopwords, performe Part of Speech (POS) tagging and Named Entity Recognition (NER) on each tweet using Stanford Named Entity Recognizer\footnote[3]{\url{https://nlp.stanford.edu/software/CRF-NER.html}} model. Using the NER tool, we extract persons, locations and organizations and removed all pseudo-documents that do not contain named entities similar to \cite{mcminn2015real}. Furthermore, We remove the tweets shorter than 3 words. 

As the Twitter maintains a maximum allowed character limit of 280 characters, collected tweets lack context information and have very low word co-occurrence. 
We tackle the challenge of topic modeling on short-text tweets and to include plentiful context information by preparing pseudo-documents for our model inputs via aggregating daily root, parent, and reply/quote/retweet comments in each conversation cascade. We maintain the order of the conversation according to the timestamps associated with each tweet. This text aggregation method results in preparing pseudo-documents rich of context and related words with a daily time resolution. 
We use the pre-processing phase output as the model input pseudo-documents, referred as documents in this paper.

\section{Conclusion and Future Directions}\label{sec:conclusion} 
In this paper, we addressed the problem of narrative modeling and narrative summary extraction for social media content. We presented a narrative framework consisting of 
\begin{enumerate*}[label=\Roman*.]
    \item Narratives over topic Categories (NOC), a probabilistic topic model with categorical time distribution; and
    \item extractive text summarization.
\end{enumerate*}
 The proposed narrative framework identifies narrative activities associated with social media events. Identifying topics' recurrence and significance over time categories with our model allowed us to propose significance-dispersity trade-off (SDT) metric. SDT can be employed as a comparison measure to identify the topic with the highest lifetime attractiveness in a timestamped corpus. 
Results on real-world timestamped data suggest that the narrative framework is effective in identifying distinct and coherent topics from the data. Additionally, the results illustrate that the identified narrative distributions approximate the user activity fluctuations over time. moreover, informative, and concise narrative summaries for timestamped data are produced. 
Further improvement of the narrative framework can be achieved via incorporating the causality relation cross the social media conversation cascades and social media events into account. Other future directions include identifying topical hierarchies and extract summaries associated with each hierarchy.

\begin{acks}
This work was supported by the Defense Advanced Research Projects Agency (DARPA) under grant number FA8650-18-C-7823. The views and opinions expressed in this article are the authors' own and should not be construed as official or as reflecting the views of the University of Central Florida, DARPA, or the U.S.
Department of Defense.
\end{acks}

\bibliographystyle{ACM-Reference-Format}
\bibliography{samples/narrative}

\vfill\eject
\appendix
\section*{Appendix}

\section{Gibbs Sampling Derivation for the Discrete-Time Narrative Model}
Starting with the joint distribution $\mathbb{P} 
(\boldsymbol{w}, \boldsymbol{t}, \boldsymbol{z}| \alpha, \beta, \psi)$, we can use conjugate priors to simplify the equations as below:

\begin{equation}
\begin{aligned}
&\mathbb{P} (\boldsymbol{w}, \boldsymbol{t}, \boldsymbol{z}| \alpha, \beta, \psi)
 = \mathbb{P}(\boldsymbol{w}| \boldsymbol{z}, \beta) \; p(\boldsymbol{t}| \psi, \boldsymbol{z}) \; \mathbb{P} (\boldsymbol{z}| \alpha) \\
& = \int \prod_{d = 1}^{D} \prod_{i = 1}^{N_d} \mathbb{P} (w_{di} | \phi_{z_{di}}) \prod_{z = 1}^{T} p(\phi_z | \beta) d\Phi \prod_{d = 1}^{D} \prod_{i = 1}^{N_d} p(t_{di} | \psi_{z_{di}}) \\
& \times \int \prod_{d = 1}^{D} \Big(\prod_{i = 1}^{N_d} \mathbb{P} (z_{di} | \theta_d) \; p(\theta_d | \alpha) \Big) d\Theta \\
&  =  \int \prod_{z = 1}^{T} \prod_{v = 1}^{V} \phi_{zv}^{n_{zv}} \prod_{z = 1}^{T} 
\Big( \frac{\Gamma (\sum_{v = 1}^{V} \beta_v)}
{\prod_{v = 1}^{V} \Gamma (\beta_v)} \prod_{v = 1}^{V} \phi_{zv}^{\beta_v-1} \Big) d\Phi \\
& \times \int \prod_{d=1}^{D} \prod_{z=1}^{T} \theta_{dz}^{m_{dz}} \prod_{d=1}^{D} 
\Big( \frac{\Gamma (\sum_{z = 1}^{T} \alpha_z)}{\prod_{z = 1}^{T} \Gamma (\alpha_z)} \prod_{z = 1}^{T} \theta_{dz}^{\alpha_z-1} \Big) d\Theta \\
& \times \prod_{d=1}^{D} \prod_{i=1}^{N_d} p(t_{di} | \psi_{z_{di}}) \\
& = \Big( \frac{\Gamma (\sum_{v = 1}^{V} \beta_v)}{\prod_{v = 1}^{V} \Gamma (\beta_v)} \Big)^{T} 
\Big( \frac{\Gamma (\sum_{z = 1}^{T} \alpha_z)}{\prod_{z = 1}^{T} \Gamma (\alpha_z)} \Big)^{D} \prod_{d = 1}^{D} \prod_{i = 1}^{N_d} p(t_{di} | \psi_{z_{di}}) \\
& \times \prod_{z = 1}^{T} \frac{\prod_{v = 1}^{V} \Gamma (n_{zv} + \beta_{v})}{\gamma (\sum_{v = 1}^{V} (n_{zv} + \beta_v))}
\prod_{d = 1}^{D} \frac{\prod_{z = 1}^{T} \Gamma (m_{dz} + \alpha_{z})}{\gamma (\sum_{z = 1}^{T} (m_{dz} + \alpha_z))},
\label{eq:apendix1}
\end{aligned}
\end{equation}
where $\mathbb{P}$ and $p$ refer to the probability mass function (PMF) and probability density function (PDF), respectively.  
The conditional probability $\mathbb{P}(z_{di}|w, t,z_{-di}, \alpha, \beta, \psi)$ can be found using the chain rule as:

\begin{equation}
\begin{aligned}
&\mathbb{P} (z_{di}|w, t,z_{-di}, \alpha, \beta, \psi) = 
\frac{\mathbb{P} (z_{di}, w_{di}, t_{di}| w_{-di}, t_{-di}, z_{-di}, \alpha, \beta, \psi)}
{\mathbb{P} (w_{di}, t_{di}|w_{-di}, t_{-di}, z_{-di}, \alpha, \beta, \psi)} \\
&\propto \frac{\mathbb{P} (w, t, z| \alpha, \beta, \psi)}{\mathbb{P} (w_{-di}, t_{-di}, z_{-di} | \alpha, \beta, \psi)} \\
&\propto \frac{n_{z_{di}w_{di}} + \beta_{w_{di}} - 1}{\sum_{v = 1}^{V} (n_{z_{di}v} + \beta_v) - 1} \;
(m_{dz_{di}} + \alpha_{z_{di}} - 1) \; p(t_{di} | \psi_{z_{di}}) \\
&\propto (m_{dz_{di}} + \alpha_{z_{di}} - 1) \; \frac{n_{z_{di}w_{di}} + \beta_{w_{di}} - 1}{\sum_{v = 1}^{V} (n_{z_{di}v} + \beta_v) - 1} \; p(t_{z_{di}} \in b_k)
\label{eq:apendix2}
\end{aligned}
\end{equation}

The probability of $p(t_{di} \in b_k)$ can be measured as follows:

\begin{equation}
\begin{aligned}
p(t_{z_{di}} \in b_k) = \frac{1}{K}\sum_{k = 1}^{K} \mathbb{I}(t_{z_{di}} \in b_k), 
\end{aligned}
\end{equation}
where $\mathbb{I}(.)$ is equal to 1 when $t_{z_{di}} \in b_k$, and 0 otherwise. 
\begin{figure*}
    \centering
    \includegraphics[width=1.2\columnwidth]{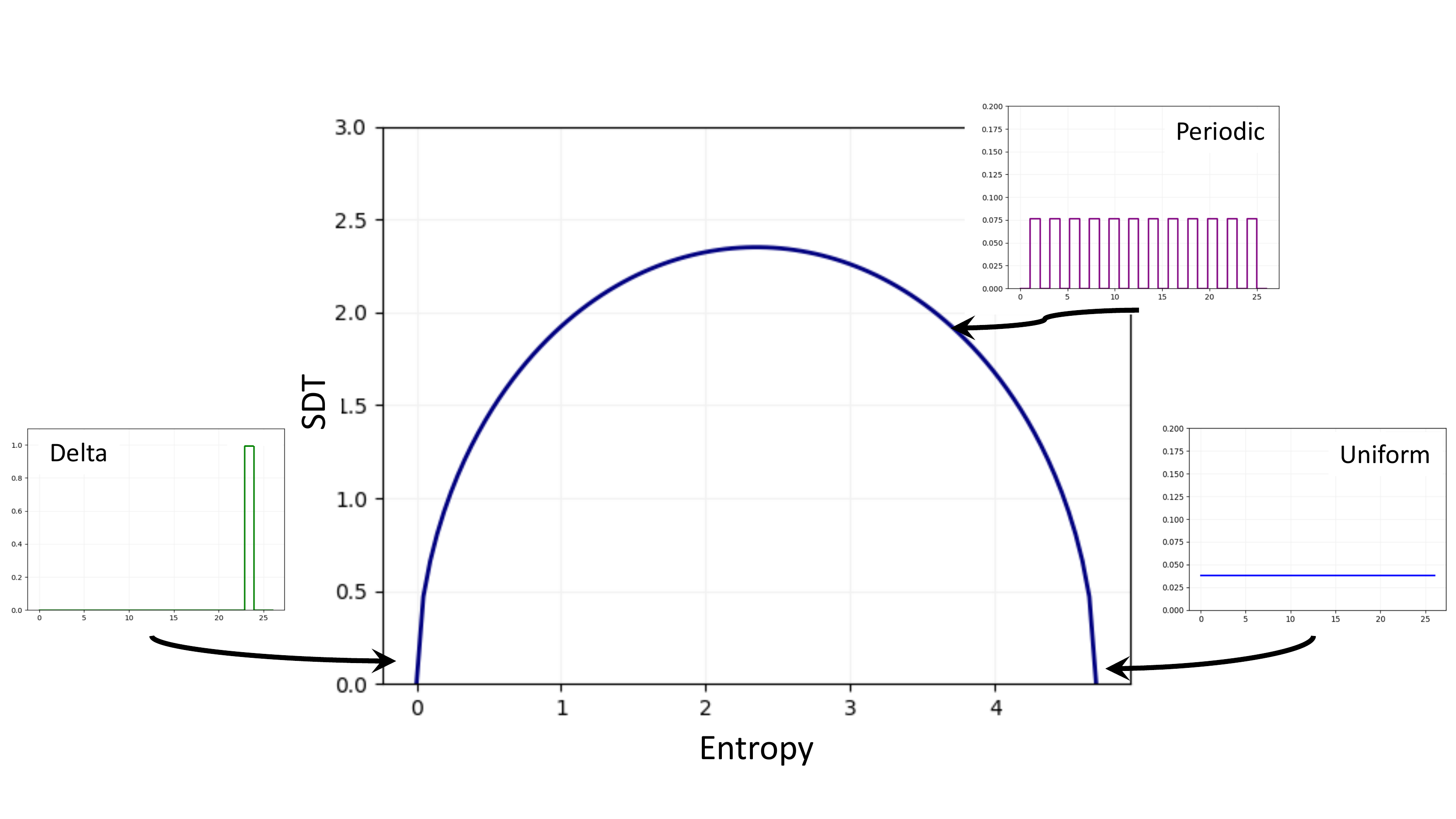}
    \caption{\small{A visualization of SDT for different entropy values using $\gamma = 0.5$. The SDT values for delta, uniform, and periodic distributions are marked on the graph.}}
    \label{fig:sdt}
\end{figure*}

\begin{table*}[h]
\caption{Representative keywords and narrative summaries per topic.}
\resizebox{\textwidth}{!}{
\begin{tabular}{ |m{1.5cm}| m{2.7cm}| m{11cm}|  }

 \rowcolor{black}
 {\color{white}{\textbf{\large{\hfil{Topics}}}}} & {\color{white}{\textbf{\large{\hfil{Keywords}}}}}	& {\color{white}{\textbf{\large{\hfil{Summary}}}}} \\
 \hline
 \textbf{\hfil{Topic 1}}	& Terrorist, Idlib, Civilian, Child, City, Attack, Aleppo, Rescue, Weapon, Killed	
 & 
 WhiteHelmets Syria News: One child was injured in the north of Aleppo. Their aim is to save lives in war zones inside Syria. 
 Has credibly substantiated 336 uses of ChemWeapons in Syria 98\% of attacks by Assadallies. These are the WHITE HELMETS or Syria Civil Defense as our US Dept of State calls them!! Russian airstrikes killed two men and one baby in DMZ areas RussianWarCrimes. \\
 \hline
 \textbf{\hfil{Topic 2}} &	Chemical, Attack, Douma, Video, Idlib, Staged, Boy, War, Child, Witness	
 & 
 Remember first they said the video including the pics of the chlorine cylinder was fake. Whitehelmets One America News Pearson Sharp Visits Hospital in Douma Where White Helmets Filmed Chemical Attack Hoax Multiple Eyewitness Doctors Say No Chemical Attack Took Place Syria. This is the video evidence of the airstrike on Zardana an Idlib town controlled by Very expensive camera on the helmet of the WhiteHelmets rescuer. 
 White Helmets making films of chemical attacks with children in Idlib. 
\\
 \hline
 \textbf{\hfil{Topic 3}}	& Chemical, Attack, Douma, Terrorist, Fake, Child, Propaganda, Video, Russian, Russia	               & 
 From the fabrication of the plays of the chemist and coverage of the crimes of terrorism to the public cooperation with the Israeli army the white helmets. 
 They are holding children! Another chemical attack is imminent its all they've got left! 4 dead including two children and more than 50 wounded mostly women and children. Love the White Helmets propaganda almost as untruthful as the BBC. 
\\
 \hline
 \textbf{\hfil{Topic 4}}	& Israel, Terrorist, Idlib, Chemical, Attack, Life, Rescue, Russian, People, Al Qaeda	               & 
WHITE HELMETS ARE PREPARING CHEMICAL ATTACK ON CITIZENS AGAIN! Those are basically just members of Al Qaeda Al Nusra right? The Al Qaeda smear is deliberate propaganda.  Its war crime only If US intervenes in Kashmir Kashmir will be liberated like Raqqan with a dozen US bases having Thaad missiles aimed at China and with AlQaeda WhiteHelmets taking out children's organs of Kashmiris. 
\\
  \hline
  \textbf{\hfil{Topic 5}}	& Funding,  Freeze, Trump, Terrorist, Group, Chemical, Attack, Idlib, Civilian, News                   & 
  Trumps USA has built a rationale for its public that it will need to support rebels in holding on to a large chunk of Syria. 
  I wonder how it is possible that criminal associations such as WhiteHelmets and the Syrian Human Rights Observatory can make the world go round as they want by influencing the policies of world leaders. U.S. freezes funding for Syrias White Helmets. White helmets are terrorists.
Former Head of Royal Navy Lord West on BBC White Helmets Aren't Neutral They're On The Side Of The Terrorists.

\\
  \hline
 \end{tabular}
 }
 \label{tab:summary}
  \begin{tablenotes}
  \item[*] \small{The summaries provided here are the results for a sample run of the proposed narrative framework and do not reflect authors' personal opinions.}
  \end{tablenotes}
 \end{table*}

\end{document}